%% file: arxiv.tex
\documentclass[12pt]{article}
\usepackage{epsfig}

\input econfmacros.tex
\def\Title#1{\begin{center} {\Large {\bf #1} } \end{center}}

\begin{document}

\Title{Contextualizing the Higgs at the LHC}

\bigskip\bigskip

%+\addtocontents{toc}{{\it D. Reggiano}}
%+\label{ReggianoStart}

\begin{raggedright}  

{\it Aleksandr Azatov, Roberto Contino, Jamison Galloway\footnote{Material presented by J.~Galloway at PLHC 2012, held at The University of British Columbia.} \\
Dipartimento di Fisica\\
Universit\`a di Roma, La Sapienza\\
{\rm and} INFN Sezione di Roma\\
I-00185 Rome, ITALY}
\bigskip\bigskip
\end{raggedright}

\begin{abstract}
Recent excesses across different search modes of the collaborations at the LHC seem to indicate the presence of a Higgs-like scalar particle at 125 GeV.  
%Operating under the assumption that these hints are indeed manifestations of a Higgs,
Using the current data sets, we review and update analyses addressing the extent to which this state is compatible with the Standard Model, and provide two contextual answers for how it might instead fit into alternative scenarios with enlarged electroweak symmetry breaking sectors.  
%These issues are addressed through application of the latest information from the ATLAS and CMS collaborations, using all publicly available data in order to constrain the properties of the 125 GeV excess as accurately as possible.
\end{abstract}

\section{Introduction}
A central issue to be addressed by the LHC is that of determining  the dynamics that UV completes the theory of massive weak vector bosons, which by itself is consistent and predictive only below scales $\mathcal O(4 \pi m/g)$, with $m$ the vector's mass and $g$ the gauge coupling.   
%Beyond this scale, the non-renormalizability of the theory requires us to introduce a new description of the physics. 

There are three possibilities for such a  completion that we will consider here:
\begin{enumerate}
\item The Higgs is an elementary state in the UV.  The nonlinear sigma model of interacting longitudinally-polarized vectors is linearized by this single state, granting the theory safe access to arbitrarily high scales in the absence of gravity.
\item The longitudinal components of the $W/Z$ states are composite, arising as Goldstone bosons of a  strong dynamics that confines at a scale $\Lambda_c \sim 4 \pi f$ with $f=246\, {\rm GeV}$, breaking a global symmetry in the pattern $G/H \sim SO(4)/SO(3)$.  There is no light Higgs boson in this minimal symmetry-breaking structure.
%A simple UV completion is accomplished by a gauge theory written in terms of new fermionic degrees of freedom above the TeV scale.
\item The longitudinal modes of the vectors are arranged into an enlarged coset space arising from a non-minimal symmetry breaking structure, e.g. $SO(5)/SO(4)$.  The Higgs can thus be realized itself as a (pseudo) Goldstone state.  
%Again the low-energy theory is completed by new gauge dynamics, with a confinement scale that can exceed that of case (2) due to the presence of a light Higgs-like particle.
\end{enumerate}

The final case above interpolates between the first two. The Higgs VEV and couplings depend on $\Lambda_c$ and a vacuum alignment angle, $\theta = \arcsin (v/f)$, determined by symmetry-breaking spurions; case (1)  corresponds to $\theta \to 0$ and case (2)  to $\theta \to 1$.  We thus find it convenient to parameterize the low-energy physics of the Higgs boson in a fully generic way, as in \cite{general}.   We restrict our study to interactions of $\mathcal O(\partial^2)$ and $\mathcal O(\partial^4)$, corresponding respectively to the first two terms in the chiral Lagrangian 
\begin{eqnarray}
\mathcal L = \mathcal L^{(2)} + \mathcal L^{(4)}+\dots
\end{eqnarray}
In unitary gauge, we have explicitly:
\small 
%\scriptsize
\begin{eqnarray}\label{eq:general}
\mathcal{L}^{(2)} &=&  \frac{1}{2} (\partial_\mu h)^2 -\frac{1}{2} m_h^2 h^2  
 - \sum_{\psi = u,d,l} m_{\psi^{(i)}} \bar \psi^{(i)} \psi ^{(i)} \left (1+c_\psi \frac{h}{v} + \dots \right) \nonumber \\
&& - \left( m_W^2 W_\mu W^\mu  + \frac{1}{2} m_Z^2 Z_\mu Z^\mu \right) \left(1 + 2a \frac{h}{v} + \dots \right);  \\
\mathcal{L}^{(4)} &=& \frac{\alpha_{em}}{4\pi}  \left(  \frac{c_{WW}}{s_W^2}\,   W_{\mu\nu}^+ W_{\mu\nu}^-  + \frac{c_{ZZ}}{s_W^2 c_W^2}\, Z_{\mu\nu}^2 + 
\frac{c_{Z\gamma}}{s_W c_W}\, Z_{\mu\nu} \gamma_{\mu\nu}   + c_{\gamma\gamma}\, \gamma_{\mu\nu}^2 \right) \frac{h}{v} \nonumber \\
&& +  \frac{\alpha_s}{4\pi} \, c_{gg}\, G_{\mu\nu}^2 \ \frac{h}{v}.
%-\frac{1}{2} (\partial_\mu h)^2 - \frac{1}{2} m_h h^2 - \frac{d_3}{6} \left( \frac{3 m_h^2}{v} \right) h^3 - \frac{d_4}{24}\left( \frac{3 m_h^2}{v^2}\right) h^4 + \dots \\
%&& - \left( m_W^2 W_\mu W^\mu  + \frac{1}{2} m_Z^2 Z_\mu Z^\mu \right) \left(1 + 2a \frac{h}{v} + b\frac{h^2}{v^2} + \dots \right)  \nonumber \\
%&&- \sum_{\psi = u,d,l} m_{\psi^{(i)}} \bar \psi^{(i)} \psi ^{(i)} \left (1+c_\psi \frac{h}{v} + c_{2 \psi} \frac{h^2}{v^2} + \dots \right). \nonumber
\end{eqnarray}
\normalsize

In what follows, we update current exclusion limits \cite{fits,us} in this parameter space and apply them first to the SM and composite Higgs scenarios with flavor-universal Yukawa rescalings ($c_\psi = c$), and second to the minimal supersymmetric Standard Model (MSSM) where up and down-type Yukawas are rescaled independently.   Explicit expressions for the overall rescaling functions of each search channel can be found for instance in the appendix of \cite{gaga}.
%We view this as a valuable preparatory exercise which can be followed through to assess the true nature of a Higgs particle, should one manifest in future data.

\section{Higgs Status: The SM and Compositeness}
We first consider the case of flavor-universal rescalings to the SM Yukawa couplings, and examine the space spanned by the couplings $a$ and $c$ of Eq.~(\ref{eq:general}).  For both ATLAS and CMS, we use best fit values for the signal strength modifier, $\mu$, to construct exclusions for all channels except those involving $h\to WW$.  For the latter, we find it more appropriate to apply the likelihood (re)construction techniques detailed in \cite{us}, which at present is the only way in which we can incorporate all subchannel information.  For the CMS searches, we reconstruct likelihoods from  the expected and observed exclusion limits provided in \cite{CMSww}, while for ATLAS we construct likelihoods using the event numbers marginalized over signal and background uncertainties quoted in \cite{ATLASww}.

\begin{figure}[htb]
\begin{center}
\epsfig{file=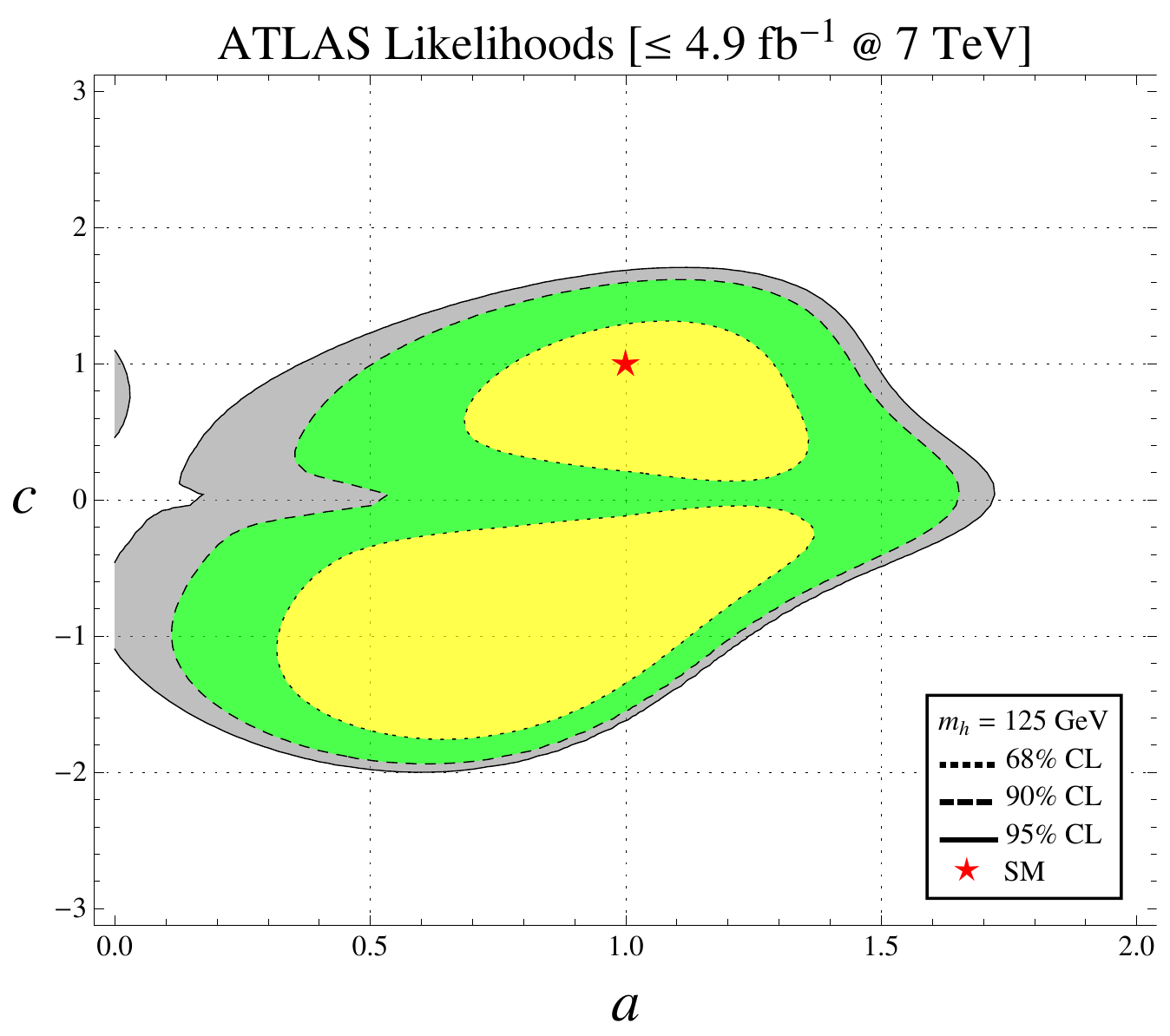,height=2.6in}
\epsfig{file=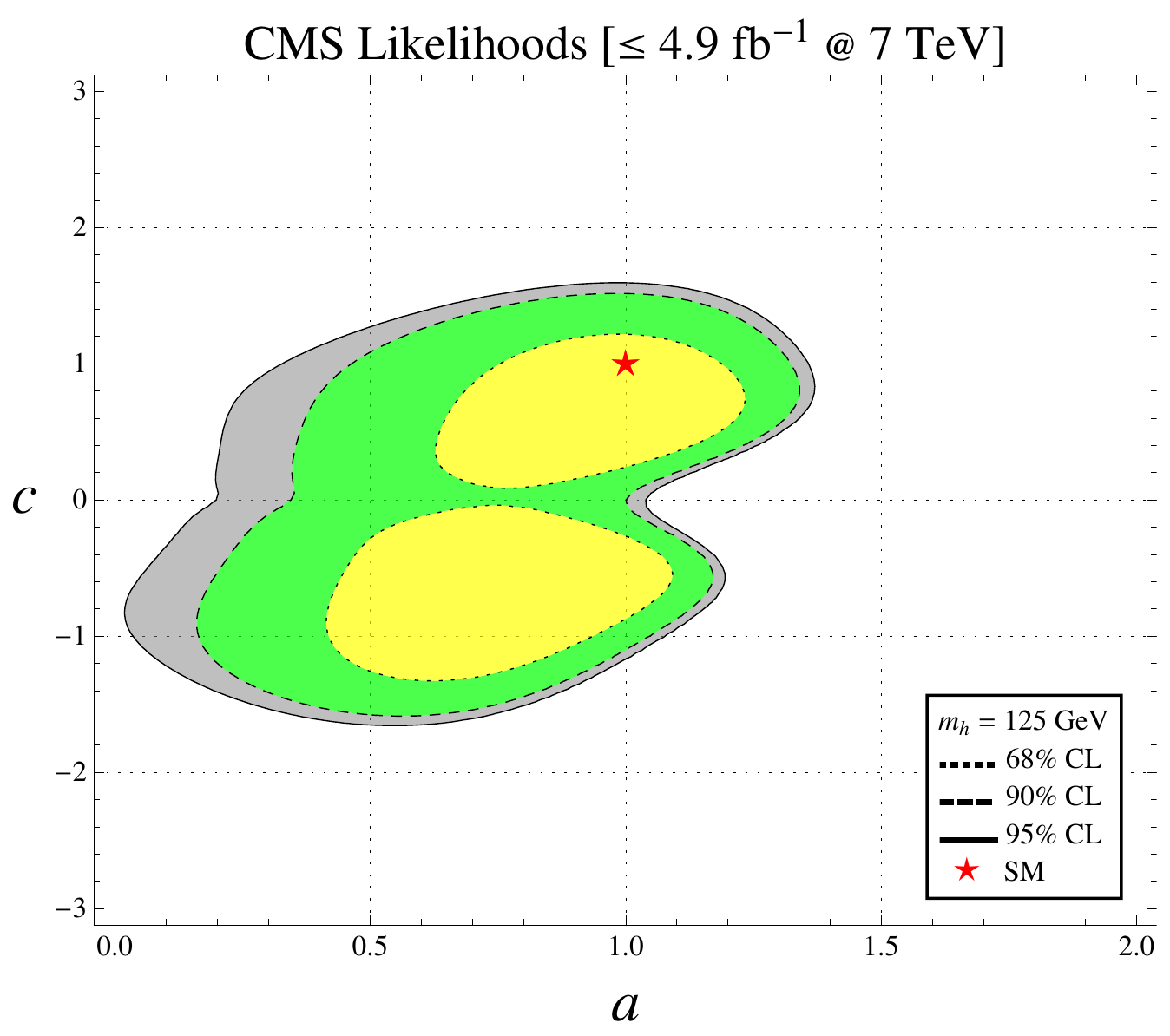,height=2.6in}
\caption{\small{Exclusion contours drawn from current ATLAS and CMS data.}}
\label{fig:ac}
\end{center}
\end{figure}

The results are shown separately for ATLAS \cite{ATLAS} and CMS \cite{CMS} in Figure~\ref{fig:ac}.  The overall picture is consistent with the SM at the $1\sigma$ level, but the peak likelihood in the neighborhood of the SM point is found to lie at $(a\simeq 0.98, c \simeq 0.58)$ with CMS data and $(a\simeq 1, c \simeq 0.73)$ with that from ATLAS.  This is suggestive of non-SM dynamics, though more data will be needed in order to draw concrete conclusions in this direction.
%allows for significant variation of the Higgs couplings even within the $1\sigma$ contour.

We note here the importance of including information of exclusive search modes, as this serves crucially to determine  even the qualitative features of the likelihood isocontours.  An important reason for the differences in constraints coming from CMS and ATLAS, for instance, can be traced to the fact that individual limits are quoted for five separate categories of the $\gamma \gamma$ \cite{CMSgaga} %and $WW$ \cite{CMSww} 
final states in CMS, while only a combined limit for this channel is quoted in the ATLAS searches.  
Treating a particular search mode at the level of exclusive subchannels should however include information regarding cut efficiencies for each category.  These can have a non-negligible impact on the effect of $\gamma \gamma$ final states in particular, and have thus been included in our fits using the estimates of \cite{gaga}.
For further details and concrete demonstrations of these effects, cf. \cite{ExcInc}.

\section{Higgs Status: Supersymmetry}
\begin{figure}[htb]
\begin{center}
\epsfig{file=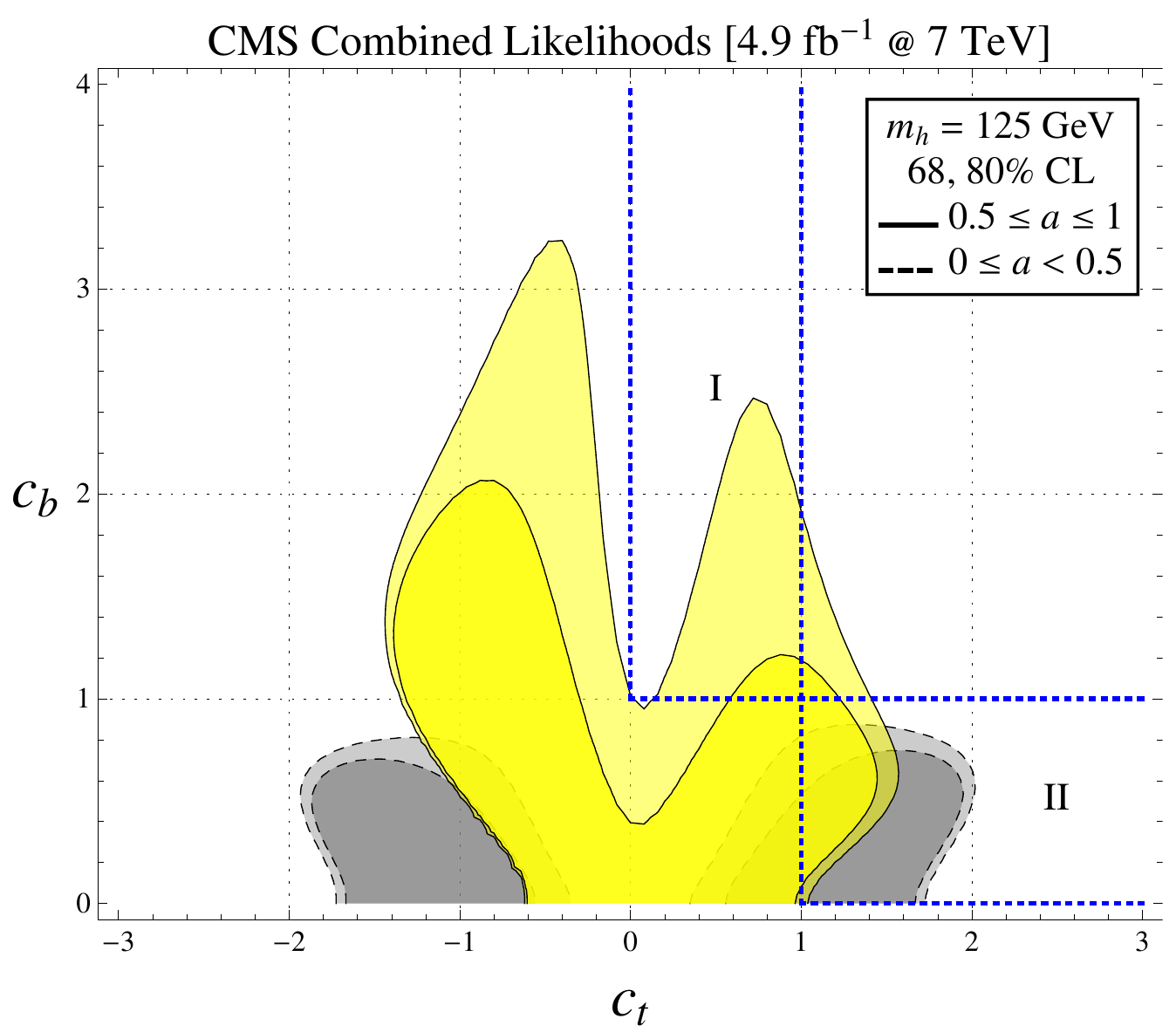,height=2.7in}
\caption{\small{3D likelihoods, projecting two ranges of vector coupling onto the plane of Yukawas for top and bottom quarks.  Regions I and II are accessible for any value of $\tan \beta$ in a generic type-II 2HDM; at tree-level in the MSSM, region II is inaccessible for any $\tan \beta < 1$.}}
\label{fig:3D}
\end{center}
\end{figure}

In the MSSM, up and down-type Yukawa couplings vary independently.  In the interest of providing  model-independent results, we show in Figure~\ref{fig:3D} likelihood contours generated in the three-dimensional space $(a, \, c_t, \, c_b=c_\tau)$ by projecting over different regions of the vector coupling $a$, with likelihoods derived by generalizing the methods described above.  We refer to \cite{Nonminimal} for further details.
%; the method of the fit we present here has been changed to match that described above, though the essential conclusions remain unchanged.

An interesting feature of Figure~\ref{fig:3D} is that for several values of the vector coupling, a preference for enhanced up-like Yukawa couplings is identified; this is not possible for the tree-level MSSM at $\tan \beta >1$, and is atypical even at loop-level with SUSY breaking effects included.  At present, the best fit point for the space accessible to the tree-level MSSM lies at the decoupling limit, where all couplings take their SM values.
%$(a\simeq 1.35,c_t\simeq 0.89,c_b\simeq1.36)$.  
This conclusion is however dominated by $h \to \gamma \gamma$, whose excesses would in fact prefer $a>1$.  The best fit within MSSM priors  is thus found at the maximal value $a=1$, with Yukawa couplings consequentially dragged to their decoupling values.  

If the $\gamma \gamma$ excesses should decrease or if there are in fact new states contributing to these decay modes such that a larger rate will continue to be observed, then it is helpful to examine separately the information from the $VV$ channels.  From these channels, we see a preference for substantial  suppression in $a$: this is due to the fact that the rescaling factor of the $WW+2j$ final state---which observes a significant reduction compared to the SM---is sensitive predominantly to $a$.  In order to accommodate this channel while maintaining more SM-like rates for the remaining inclusive $VV$ modes, the decrease in production via vector boson fusion due to $a<1$ is offset by an increase in the top contribution to gluon fusion, i.e. $c_t>1$ and thus $c_b <1$.

%%%%%%%%%%%%%%%%%%%%%%%%%%%%%%%%%%%%%%%%%%%%%%%%%%%%%%%%%%%%%%%%%%%%%%%%%
%%
%%   use this format to include an .eps figure into your paper
%%
%%%%%%%%%%%%%%%%%%%%%%%%%%%%%%%%%%%%%%%%%%%%%%%%%%%%%%%%%%%%%%%%%%%%%%%%%%%

%%%%%%%%%%%%%%%%%%%%%%%%%%%%%%%%%%%%%%%%%%%%%%%%%%%%%%%%%%%%%%%%%%%%%%%%%
%%
%%   use this format to include a LaTeX table  into your paper
%%
%\begin{table}[b]
%\begin{center}
%\begin{tabular}{l|ccc}  
%Patient &  Initial level($\mu$g/cc) &  w. Magnet &  
%w. Magnet and Sound \\ \hline
% Guglielmo B.  &   0.12     &     0.10      &     0.001  \\
% Ferrando di N. &  0.15     &     0.11      &  $< 0.0005$ \\ \hline
%\end{tabular}
%\caption{Blood cyanide levels for the two patients.}
%\label{tab:blood}
%\end{center}
%\end{table}
%%%%%%%%%%%%%%%%%%%%%%%%%%%%%%%%%%%%%%%%%%%%%%%%%%%%%%%%%%%%%%%%%%%%%%%%%%%

\section{Conclusions}
The current excesses observed by the ATLAS and CMS collaborations are found to be consistent with the SM taking all data combined, though room for significant deviations obviously remains.  The statistics of the available data is still sufficiently low that we expect these conclusions to remain very much in flux during the coming months.  In this note, we have reviewed ways in which model-independent fits can be constructed in generic parameter spaces, and provided contextual answers for whether the current situation allows for the hinted Higgs-like state to emerge from a larger EWSB sector. The likelihoods in the context of flavor-universal composite Higgs peak at suppressed Yukawa couplings which could be indicative of relatively low-scale compositeness, while an interesting shallow direction exists in the MSSM fit which could cause tension for minimality and therefore suggest some other new SUSY dynamics to be relevant for weak-scale physics.
\\
$$
* \ * \ * 
$$
\\
J.G. thanks the organizers of PLHC 2012 for the invitation to present this material, and is grateful for many stimulating discussions in the very hospitable environment of The University of British Columbia.

%\bigskip
%I am grateful to Don Alfonso d'Alba for certain services essential to 
%this investigation.

\def\Discussion{
\setlength{\parskip}{0.3cm}\setlength{\parindent}{0.0cm}
     \bigskip\bigskip      {\Large {\bf Discussion}} \bigskip}
\def\speaker#1{{\bf #1:}\ }
\def\endDiscussion{}

%\Discussion

%\speaker{D. Giovanni (University of Seville)}  My analysis indicates that the
%recovery of the two gentlemen is due simply to their embrace of the masculine
%principle and has nothing to do with magnetism at all.  Could you comment on 
%this?

%\speaker{Reggiano} Professor Giovanni has discussed this hypothesis in several
%forums, but, I do not believe there is anything in print.  I understand that
%he is spending his time in other pursuits.

%\speaker{D. Anna (University of Seville)}  In fact, my colleague Giovanni 
%has expressed opposite opinions on this question at various times, depending
%on the audience.  All of these testosterone-based theories are, of course,
%nonsense.

%\endDiscussion
 
\end{document}

%% file: econfmacros.tex
\def\beq{\begin{equation}}
\def\eeq#1{\label{#1}\end{equation}}
\def\eeqn{\end{equation}}

\def\beqa{\begin{eqnarray}}
\def\eeqa#1{\label{#1}\end{eqnarray}}
\def\eeqan{\end{eqnarray}}

\let\bar=\overbar

\def\Dslash{\not{\hbox{\kern-4pt $D$}}}
\def\dslash{\not{\hbox{\kern-2pt $\del$}}}

\def\msb{{\bar{\ssstyle M \kern -1pt S}}}